\documentclass{kluwer}
\usepackage[]{graphicx}
\begin{document}
\begin{article}
\begin{opening}
\title{The SCUBA Local Universe Galaxy Survey.}            

\author{Loretta \surname{Dunne}\email{L.Dunne@astro.cf.ac.uk}}
\author{Stephen A. \surname{Eales}} 
\runningauthor{L. Dunne}
\runningtitle{The SCUBA Local Universe Galaxy Survey}
\institute{Cardiff University}                               


\end{opening}

The SCUBA Local Universe Galaxy Survey (SLUGS) is the first systematic
survey of the local universe at submm wavelengths. At 850$\mu$m, SCUBA
is sensitive to thermal emission from dust at $\gsim 10$ K and should
therefore trace the bulk dust mass, unlike IRAS which was only
sensitive to warmer dust ($>30$ K) near star-forming regions. So far,
we have imaged $\sim 200$ objects selected from the IRAS Bright Galaxy
Sample (Soifer et al. 1989) and the CfA optical redshift survey
(Huchra et al. 1983), and first estimates of the 850$\mu$m luminosity
and dust mass functions have been produced from the IRAS sample (Dunne
et al. 2000). This luminosity function may be biased if a population
of submm emitting galaxies exists which are not present in our
60$\mu$m flux limited survey. Such a population could consist of
galaxies with large amounts of cold dust ($<25 $ K), which would be
strong 850$\mu$m sources but very weak at 60$\mu$m. We have used
450$\mu$m data for 32 bright IRAS galaxies to investigate the presence
of cold dust components, and we find that they do exist at $\sim 20$ K
in even the most luminous IRAS objects (Dunne \& Eales 2001). The
selection of our optical sample should not be affected by dust
temperature, and so provides a way to address the question of whether
a `cold' population, whose FIR emission is dominated by this
$\leq 20$ K component indeed exists.

\begin{figure}
\centerline{\includegraphics[width=10pc,angle=-90]{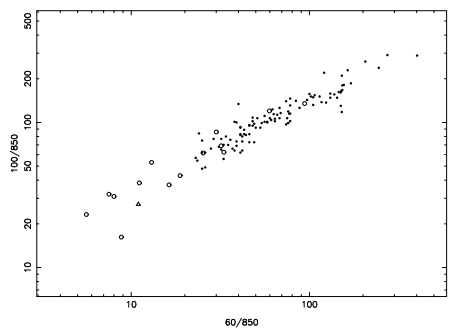}}
\caption{IRAS/submm colour plot for the IRAS (filled
dots) and optically selected objects (open circles). Some of the
optically selected galaxies overlap with the IRAS sample in terms of
their colours, but many show very much `colder' colours and occupy the
bottom left region of the plot where no IRAS selected objects are
found.}\label{coloptF}
\end{figure}

\begin{figure}
\tabcapfont
\centerline{%
\begin{tabular}{c@{\hspace{2pc}}c}
\includegraphics[width=1.6in,angle=-90]{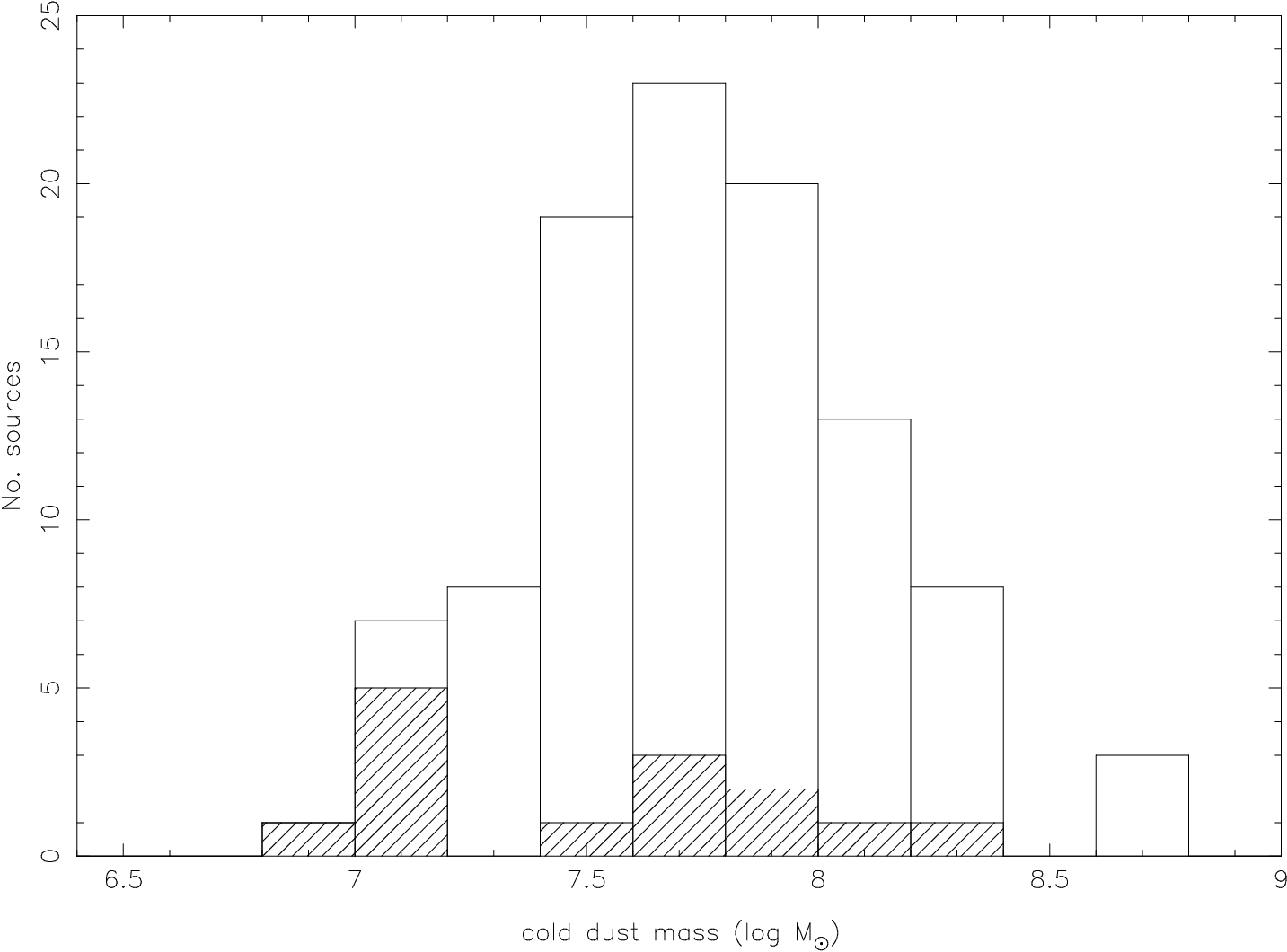}&
\includegraphics[width=1.6in,angle=-90]{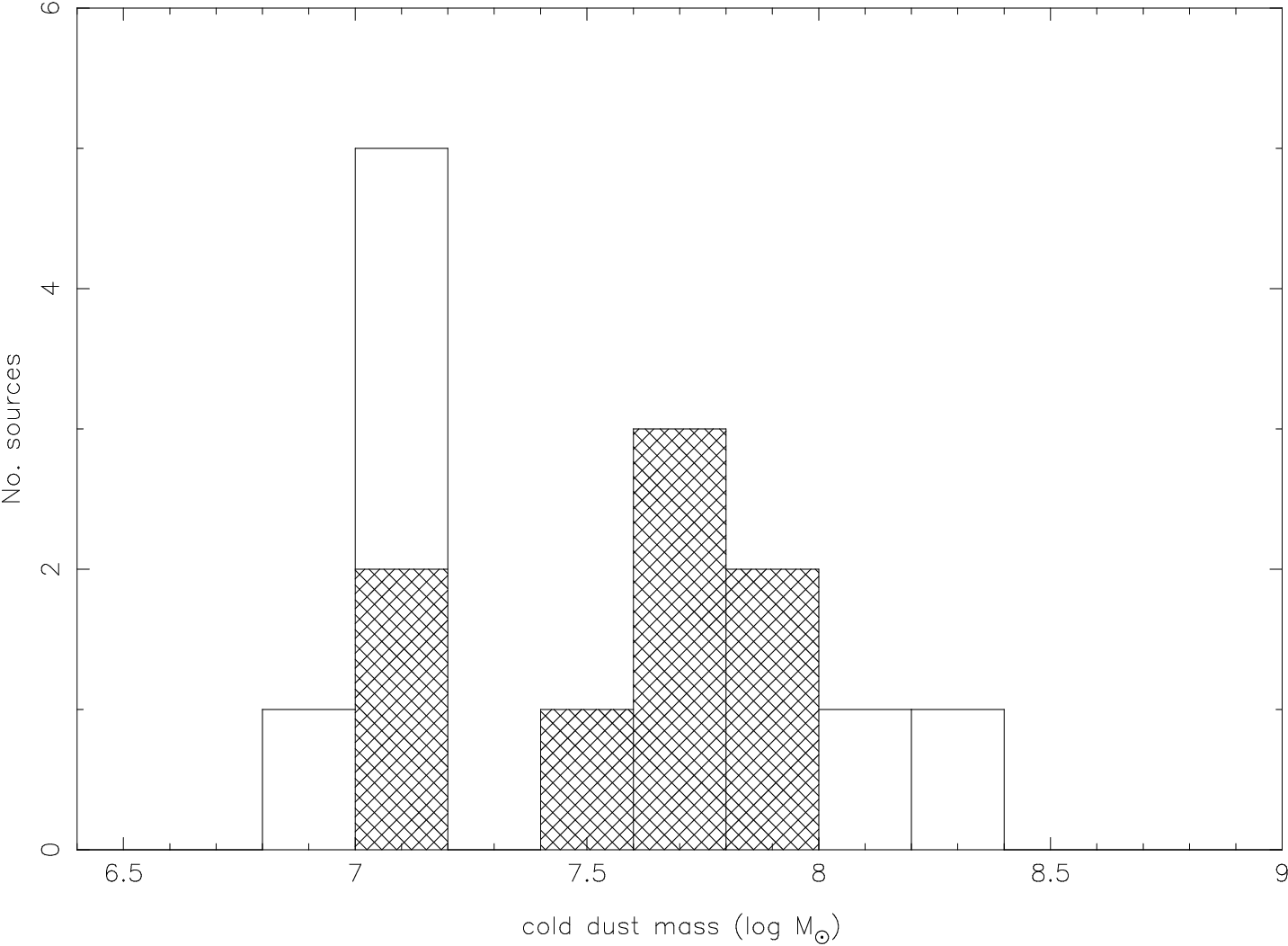}\\
a. dust masses (IRAS and optical)  & b. dust masses (optical only)
\end{tabular}}
\caption{Distribution of (a) IRAS and optically (shaded) selected dust
masses (b) dust masses for the optically selected galaxies as a function of
IRAS/submm colour. Shaded -- `cold' optically
selected galaxies with very low IRAS/submm colours.}\label{mdhistF}
\end{figure}

Fig.~\ref{coloptF} compares the IRAS/submm colours for the IRAS BGS
sample and some of the optically selected galaxies for which we have
reduced the submm data. Around half of the optically selected galaxies
lie outside the range of colours defined by the IRAS sample. Clearly,
this data does not support the scenario where all galaxies have a
similar colour, with the same objects detected in both 60$\mu$m and
850$\mu$m surveys. For our purposes, a `bias' is present if those
optical galaxies not in the IRAS region of Fig.~\ref{coloptF} would
have been detected in a SCUBA blind survey to the same depth, or put
simply, if the dust masses and $L_{850}$ of these `cold' objects are within
the distribution for the IRAS sample. Fig.~\ref{mdhistF}a shows the
distribution of dust masses for the IRAS and optically selected
galaxies. The optically selected have a lower mean dust mass but there
is still a substantial overlap with the IRAS
sample. Fig.~\ref{mdhistF}b shows the optically selected dust masses
separated into those objects with very `cold' IRAS/submm colours and
those which are similar to the IRAS galaxies. The `cold' objects make
up the middle of the distribution, and from comparison to (a) have
masses similar to the median of the IRAS galaxies. These provisional
results suggest that a population of objects dominated by cold
components does exist and that, in terms of dust mass, these galaxies
are comparable to those selected by IRAS. The optically selected
sample is therefore essential for providing a complete view of the
local submm universe and for addressing the bias in the LF and dust
mass functions presented in Dunne et al. 2000.

{}
\end{article}

\begin{thebibliography}{}
\bibitem[]{} Dunne, L., et al.: 2000, {\em MNRAS\/} {\bf 115}, pp.315
\bibitem[]{} Dunne, L. and Eales, S.A.: 2001, {\em MNRAS\/} {\bf 327},
pp. 697
\bibitem[]{} Huchra, J., Davis, M., Latham, D., Tonry, J.: 1983, {\em
ApJS\/} {\bf 52}. pp. 89
\bibitem[]{} Soifer, B.T., Boehmer, L., Neugebauer, G., Sanders, D.B.:
1989, {\em AJ\/} {\bf 98}, pp. 766
\end{thebibliography}
\end{document}